\documentclass[journal]{vgtc}                

\ifpdf
  \pdfoutput=1\relax                   
  \usepackage{graphicx}                
  \DeclareGraphicsExtensions{.pdf,.png,.jpg,.jpeg} 
\else
  \usepackage{graphicx}                
  \DeclareGraphicsExtensions{.eps}     
\fi%
\usepackage{color}
\graphicspath{{figures/}{pictures/}{images/}{./}} 

\usepackage{microtype}                 
\PassOptionsToPackage{warn}{textcomp}  
\usepackage{textcomp}                  
\usepackage{mathptmx}                  
\usepackage{times}                     
\usepackage{cite}                      
\usepackage{tabu}                      
\usepackage{booktabs}                  
\usepackage{enumitem}
\usepackage{amsmath}
\usepackage{setspace}
\usepackage{amsfonts}

\newcommand{\tang}[1]{{\textcolor{black}{#1}}}

\onlineid{1091}

\vgtccategory{Research}
\vgtcpapertype{Technique}

\title{PlotThread: Creating Expressive Storyline Visualizations using Reinforcement Learning}

\author{Tan Tang, Renzhong Li, Xinke Wu, Shuhan Liu, Johannes Knittel, Steffen Koch, \\ Thomas Ertl, Lingyun Yu, Peiran Ren, and Yingcai Wu}
\authorfooter{
\item
 T. Tang, R. Li, X. Wu, S. Liu, Y. Wu are with Zhejiang Lab and State Key Lab of CAD\&CG, Zhejiang University. E-mail: {tangtan, renzhongli, xinke\_wu, shliu, ycwu}@zju.edu.cn.
 Y. Wu is the corresponding author.
\item
 J. Knittel, S. Koch, and T. Ertl are with VIS/VISUS, University of Stuttgart. E-mail: {Johannes.Knittel, Steffen.Koch, Thomas.Ertl}@vis.uni-stuttgart.de.
\item
 L. Yu is with Department of Computer Science and Software Engineering, Xi'an Jiaotong-Liverpool University. Email: Lingyun.Yu@xjtlu.edu.cn.
\item
 P. Ren is with Alibaba Group. E-mail: renpeiran@gmail.com.
}

\shortauthortitle{Biv \MakeLowercase{\textit{et al.}}: Global Illumination for Fun and Profit}

\abstract{
  Storyline visualizations are an effective means to present the evolution of plots and reveal the scenic interactions among characters.
  However, the design of storyline visualizations is a difficult task as users need to balance between aesthetic goals and narrative constraints.
  Despite that the optimization-based methods have been improved significantly in terms of producing aesthetic and legible layouts, the existing (semi-) automatic methods are still limited regarding 1) efficient exploration of the storyline design space and 2) flexible customization of storyline layouts.
  In this work, we propose a reinforcement learning framework to train an AI agent that assists users in exploring the design space efficiently and generating well-optimized storylines.
  Based on the framework, we introduce PlotThread, an authoring tool that integrates a set of flexible interactions to support easy customization of storyline visualizations.
  To seamlessly integrate the AI agent into the authoring process, we employ a mixed-initiative approach where both the agent and designers work on the same canvas to boost the collaborative design of storylines.
  We evaluate the reinforcement learning model through qualitative and quantitative experiments and demonstrate the usage of PlotThread using a collection of use cases.
} 


\CCScatlist{ 
 \CCScat{K.6.1}{Management of Computing and Information Systems}%
{Project and People Management}{Life Cycle};
 \CCScat{K.7.m}{The Computing Profession}{Miscellaneous}{Ethics}
}

\teaser{
  \centering
  \includegraphics[width=0.9\linewidth]{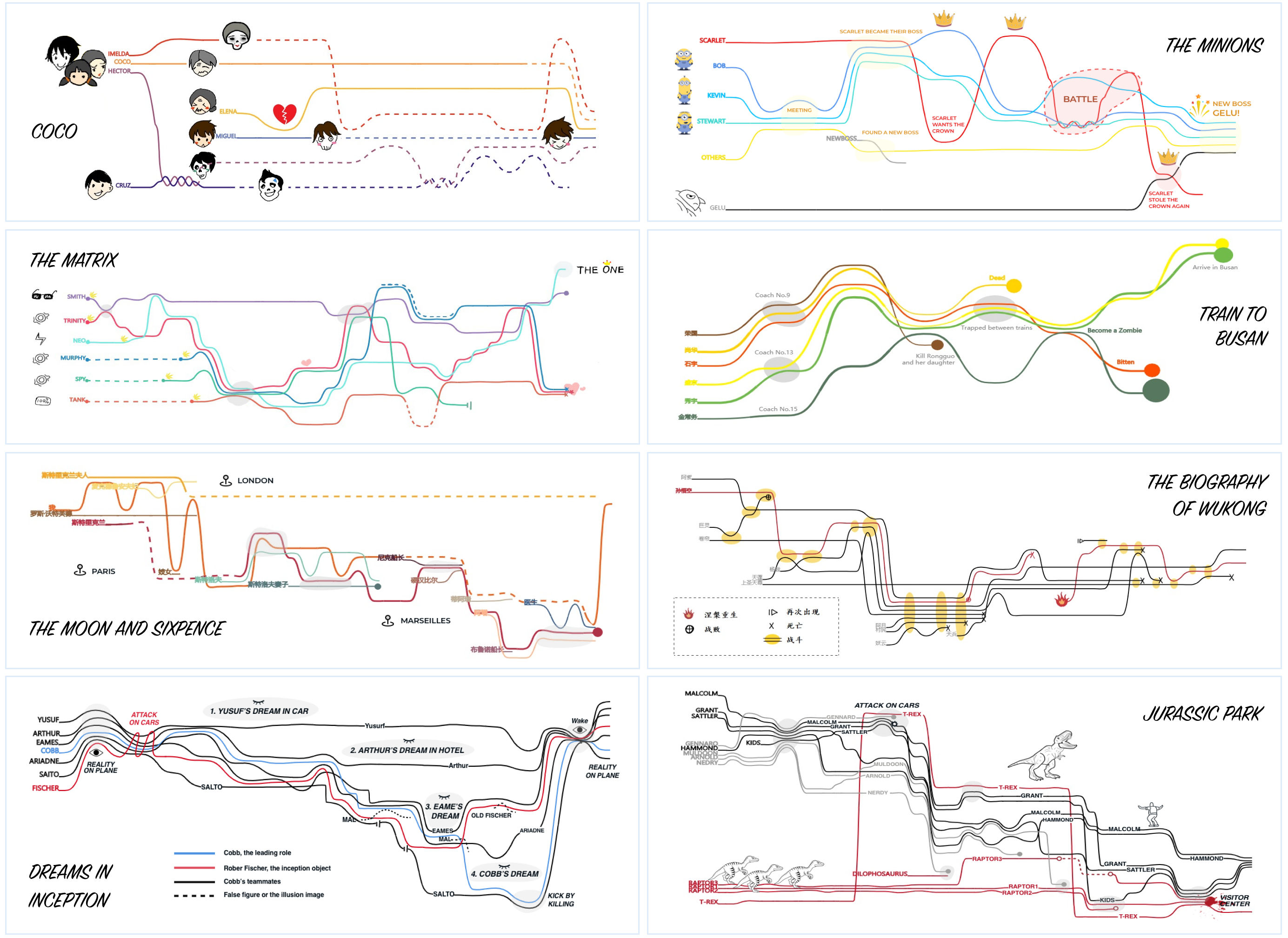}
  \caption{Example storyline visualizations created using PlotThread.
  The layouts are generated through a collaborative design between the AI agent and designers, while the styles and visual labels are customized manually to embellish the storylines.
  }
	\label{fig:teaser}
}

\vgtcinsertpkg

\begin{document}


\firstsection{Introduction}

\maketitle

\begin{spacing}{0.98}

Storyline visualizations~\cite{xkcd,Tanahashi2012} have gained wide popularity in presenting complex entity relationships.
The ability to create visual narratives~\cite{Lee2013} makes it applicable in presenting fictions~\cite{xkcd}, analyzing dynamic networks~\cite{Liu2013}, recalling meeting content~\cite{Shi2018a}, and understanding software evolutions~\cite{Ogawa2010}.
However, designing storyline visualizations has long been considered as a difficult and tedious task which involves balancing the trade-off between narrative constraints~\cite{Tang2018a} and aesthetic goals~\cite{Liu2013}.
To illustrate the evolutions of entity relationships, two primary narrative constraints~\cite{Tang2018a} should be followed:
\begin{itemize}[nosep]
  \item \textbf{C1} \textit{the lines that represent characters who appear in the same scene should be grouped.}
  \item \textbf{C2} \textit{otherwise, the grouped lines should be divided.}
\end{itemize}

Inspired by graph layouts~\cite{Tamassia1988}, it is necessary to minimize line crossings and deviations to avoid dense visual clutter.
Thus, three aesthetic goals~\cite{Tanahashi2012} are proposed to create legible layouts:
\begin{itemize}[nosep]
  \item \textbf{G1} \textit{reducing line crossings}
  \item \textbf{G2} \textit{reducing line wiggles}
  \item \textbf{G3} \textit{reducing white space}
\end{itemize}

To ease the difficulty of designing storyline visualizations, previous studies~\cite{Arendt2017,Liu2013,Tanahashi2012,Tanahashi2015} have developed optimization-based methods that produce storylines according to the design factors mentioned above.
However, these methods mainly focus on producing aesthetic and legible layouts without considering the whole design space of storylines.
With limited design choices, the storylines generated by the optimization models cannot cover the diverse narrative elements compared to the manually-created ones~\cite{Tang2018a}.
For example, the hand-drawn storylines~\cite{xkcd,Tang2018a} adopt various layouts to indicate different plots.

To support the design of expressive storylines from the narrative aspect, researchers~\cite{Tang2018a} developed iStoryline that incorporates human design knowledge and creativity into the optimization models.
Specifically, iStoryline formulates user interactions as mathematical constraints to control the optimization model~\cite{Liu2013} so that users can focus on constructing storyline layouts that conform to their understandings of the stories.
However, the interactions proposed in iStoryline mainly concentrate on modifying the local regions, which makes it time-consuming and labor-intensive to customize the overall layouts.
For example, users may need to take a considerable number of actions to refine storyline layouts, which hinders the efficient exploration of the design space.
Besides, the unpredictability of the optimization process may give rise to unexpected results, which requires trial-and-error practices to obtain the desired storylines.

To facilitate the easy design of storyline layouts, we envision whether a human-AI (\textbf{A}rtificial \textbf{I}ntelligence) collaborative approach can be helpful.
Specifically, we intend to employ machine learning to develop an intelligent agent.
Similar to a recommendation engine, the agent can reduce human efforts by providing users possible suggestions of compelling storylines that follow the aesthetic goals (\textbf{G1} to \textbf{G3}).
However, we are not aware of any prior work on designing storylines using machine learning, which raises two major challenges:

\textbf{Model Architecture}
Storylines depict the temporal relationships~\cite{xkcd,Tanahashi2012} among entities that are inherently different from the Euclidean data (e.g., images, videos, and texts) that can be processed by existing machine learning applications~\cite{Kwon2019a,Wang2019a}.
Thus, it remains unclear whether storylines can be generated using machine learning or how to extend the existing models to deal with storylines.
Recent studies~\cite{Wang2019a,Kwon2019a} have adapted neural networks for graph drawings, but they mainly focus on the topological structure of graph layouts.
While storylines and graphs pursue some common aesthetic goals (e.g., \textit{minimizing crossings}~\cite{Diaz2002}), storylines require a higher aesthetic standard for legible layouts.
Moreover, it is also necessary to develop a novel learning framework that takes narrative constraints into considerations for the storyline generation problem.

\textbf{Model Training}
Training a machine learning model requires an appropriate loss function and a high-quality benchmark dataset~\cite{He2016a}.
In image classification, for instance, the loss function can be easily defined as counting incorrect labels while the training data can be obtained by labeling real-world images~\cite{Huang2019a}.
However, the training of the storyline model becomes more complicated than typical machine learning tasks.
First, it is challenging to define ``correct'' layouts in terms of the different narratives since designers usually have different understandings about the stories.
Thus, it is difficult to identify a unique loss function for the storyline generation problem.
Second, there are not enough storyline visualizations available to train a machine learning model, even though previous work~\cite{Tang2018a} extended the collection of hand-drawn storylines.

In this work, we propose a novel reinforcement learning framework that trains an AI agent to design storylines ``like'' human designers.
To support the collaborative design, the agent should follow two principles:
\begin{itemize}[nosep]
  \item \textbf{D1} \textit{Storylines generated by agents should resemble the ones on which users are currently working to preserve their mental map.}
  \item \textbf{D2} \textit{The agent should share the same action space as human users so that they can work on the same canvas collaboratively.}
\end{itemize}

Thus, the goal of the AI agent is to imitate and improve users' intermediate results instead of generating storylines from scratch.
To achieve this goal, the agent should be capable of decomposing a given storyline into a sequence of actions, understanding the state of intermediate layouts, and have a foresightful plan for future actions.
Therefore, we employ \textbf{R}einforcement \textbf{L}earning (RL) to solve the challenges.
Specifically, we define the \textit{states} as the intermediate storyline layouts and define the \textit{actions} of the agent as the same interactions implemented by iStoryline due to its success in producing diverse storylines that conform to different narratives.
We further define loss function by maximizing the accumulative \textit{rewards} that are vital for training RL models.
To obtain sufficient training data, we follow the common practices~\cite{Wang2019a,Kwon2019a} that generate well-optimized storylines with diverse visual layouts using the existing optimization approach~\cite{Liu2013}.

As a proof of concept, we implement PlotThread that integrates the agent into the authoring process of storyline visualizations.
We extend the interaction set of iStoryline to support a more flexible design of storylines and foster close collaboration between the agent and designers.
We present the usage of PlotThread through a set of use cases (see Fig.~\ref{fig:teaser}) and validate its usability via expert interviews.

The main contributions are summarized as follows:
\begin{itemize}[nosep]
  \item We propose a novel reinforcement learning framework and generate a collection of high-quality storylines to train an agent that supports the collaborative design of storylines with designers.
  \item We develop PlotThread, a mixed-initiative system that facilitates the easy creation of expressive storyline visualizations, and demonstrates its usage through a set of use cases.
\end{itemize}

\section{Related Work}
We summarize critical techniques used in producing storyline visualizations and the state-of-the-art reinforcement learning techniques.

\subsection{Storyline Visualization} 
Storyline visualizations have become prevalent in revealing the evolution of stories~\cite{Liu2013} and presenting various narrative elements~\cite{Tang2018a}.
To ease the difficulties in designing storyline layouts, researchers have proposed many (semi-) automatic approaches~\cite{Tanahashi2012,Tanahashi2015,Liu2013,Arendt2017} that achieve the trade-off between aesthetic goals and narrative constraints using optimization models.
Ogawa and Ma~\cite{Ogawa2010} firstly proposed an automatic approach for generating storyline visualizations but their algorithm failed to produce aesthetic layouts due to the ignorance of the heuristic criteria.
Tanahashi and Ma~\cite{Tanahashi2012} suggested a more comprehensive set of design considerations for storyline visualizations and proposed a layout generation approach based on genetic algorithms.
Despite the success of producing relatively aesthetically-appealing and legible storyline layouts, their technique is inefficient to support interactive editing of storyline visualizations.
For a better performance in both efficiency and the overall aesthetic quality, StoryFlow~\cite{Liu2013} was developed to generate storyline visualizations using a hybrid approach that combines discrete and continuous optimization models.
Moreover, it supports real-time interactions (e.g., bundling, removing, straightening) for users to edit storyline layouts.
However, the automatically-generated storylines are not comparable to the hand-drawn illustrations~\cite{Tang2018a} in terms of the expressiveness because the automatic methods cannot cover abundant narrative elements, including plots, tones, etc.

To create more meaningful storyline visualizations that conform to designers' requirements, Tang et al.~\cite{Tang2018a} extended the design space of storylines that associates narrative elements with visual elements.
They further developed iStoryline that integrates a set of high-level post-editing interactions to support the flexible customization of storyline layouts.
They developed a set of easy-to-use high-level interactions, but it is still inefficient to explore the design space and construct the overall layout using these fine-grained interactions.
iStoryline automatically translates the high-level interactions into mathematical constraints which are further integrated into the optimization model~\cite{Liu2013} to generate storyline layouts.
However, users may obtain unexpected layouts due to the unpredictability of the optimization process, which requires trial-and-error practices to refine the results.
To improve user experiences, we employ reinforcement learning to reduce users' effort in iteratively refining storyline visualizations.

\subsection{Reinforcement Learning}

Reinforcement learning refers to a system where an \textit{agent} performs a task using a set of \textit{actions} in an \textit{environment} that can be represented by a set of \textit{states}~\cite{Kaelbling1996a,yuan2021survey}.
The learning process can be depicted by an agent predicting the ``next'' action based on the observed ``current'' state and obtain a \textit{reward}~\cite{Sutton2018a}, and the goal of the agent is to maximize cumulative \textit{rewards}.
Due to the emergent development of deep learning techniques~\cite{He2016a,Schmidhuber2015a}, deep reinforcement learning~\cite{Yang2017a,Racaniere2017a} (DRL) has burgeoned in fields like games~\cite{Kempka2016a,Mnih2015a} and painting~\cite{Huang2019a}.
Mnih et al.~\cite{Mnih2015a} proposed a deep Q-network to perform a group of challenging tasks in classic 2D games for the Atari 2600 console~\cite{Bellemare2012a} and achieved great success in surpassing the previous algorithms when performing the same tasks.
To simulate a semi-realistic 3D world, Kempaka et al.~\cite{Kempka2016a} introduced a new AI research platform called ViZDoom and further employed deep Q-learning and experience replay to train competent agents.
To demonstrate how to teach machines to paint like human painters, Huang et al.~\cite{Huang2019a} employed a neural renderer in model-based DRL to train an agent that creates fancy drawings using a small group of strokes.
Despite that reinforcement learning has become prevalent in various fields, we are not aware of any prior works on designing storylines.
The issue of producing storylines is similar to the graph drawing problem~\cite{Wang2019a,Kwon2019a} because their ultimate goal is to obtain well-designed layouts.
To achieve this goal, Wang et al.~\cite{Wang2019a} employ a graph-LSTM-based model to map graph structures to graph layouts directly, and Kwon et al.~\cite{Kwon2019a} employ a deep generative model that uses an encoder-decoder framework to map training datasets into a latent space.
However, the existing approaches are not applicable to our work because storylines pursue higher aesthetic criteria~\cite{Wang2005a} and need to balance narrative constraints~\cite{Tang2018a}.
Thus, we intend to develop a novel reinforcement learning framework that trains an AI agent to design storylines like human users to support collaborative design.

\section{PlotThread}
We develop a mixed-initiative tool~\cite{Stefnisson2017a}, PlotThread, to facilitate the easy creation of storyline visualizations.
We believe it is essential to combine both human and AI intelligence so that designers can produce creative storylines based on their design preferences and understandings about stories while the agent can reduce labor-intensive efforts.

\subsection{Design Considerations}
The mixed-initiative application~\cite{Liapis2013a} refers to a system where automated services (e.g., agents) and users work iteratively (i.e., taking turns) to perform tasks in the same context~\cite{Liapis2016a,Novick1997a}.
Design principles for mixed-initiative systems have been explored~\cite{Horvitz1999a,Horvitz1999b} to achieve effective collaboration between users and computers.
To guide the design of PlotThread, we summarize two primary design considerations:

\vspace{2mm}
\noindent
\textbf{DC1. Support a smooth collaborative design workflow.}
The AI agents could act as a stimulus for lateral thinking~\cite{De2010a} to inspire co-creativity~\cite{Yannakakis2014a}.
To foster effective human-AI collaboration, it is necessary to place the human at the center of visualization designs, while the AI agent should assist, rather than replace the designers~\cite{Yannakakis2014a}.
Hence, users should be granted enough control in the decision-making stage.
One common practice is that the user takes the task-initiate~\cite{Novick1997a} in customizing an initial layout.
Then, the agent proactively contributes to the design process by improving users' intermediate results and providing alternative designs based on users' input layouts.
Moreover, users should be capable of further modifying and improving the AI-generated storyline instead of merely accepting or rejecting it.
To follow this practice, it is essential to seamlessly integrate the AI agent into the authoring process and provide a smooth co-design workflow.

\vspace{2mm}
\noindent
\textbf{DC2. Balance fine-grained and high-level interactions.}
It is burdensome for users to create storyline layouts while pursuing the aesthetic goals (\textbf{G1} to \textbf{G3}).
For reducing human efforts, the previous study~\cite{Tang2018a} proposed high-level interactions that can invoke the optimization model to re-layout storylines.
The high-level interactions enable users to change the overall layouts while the aesthetic quality is ensured by the optimization model~\cite{Liu2013}.
While they are easy to use, the high-level interactions cannot fully support users' design requirements due to their limited flexibility.
Conversely, fine-grained interactions focus on modifying the individual lines, so they are flexible enough to support various design requirements.
However, they are also tedious and even require professional skills.
Since ``users may often wish to complete or refine an analysis provided by an agent''~\cite{Horvitz1999a}, we need to achieve a better balance between the high-level and fine-grained interactions.

\begin{figure}
  \includegraphics[width=0.5\textwidth]{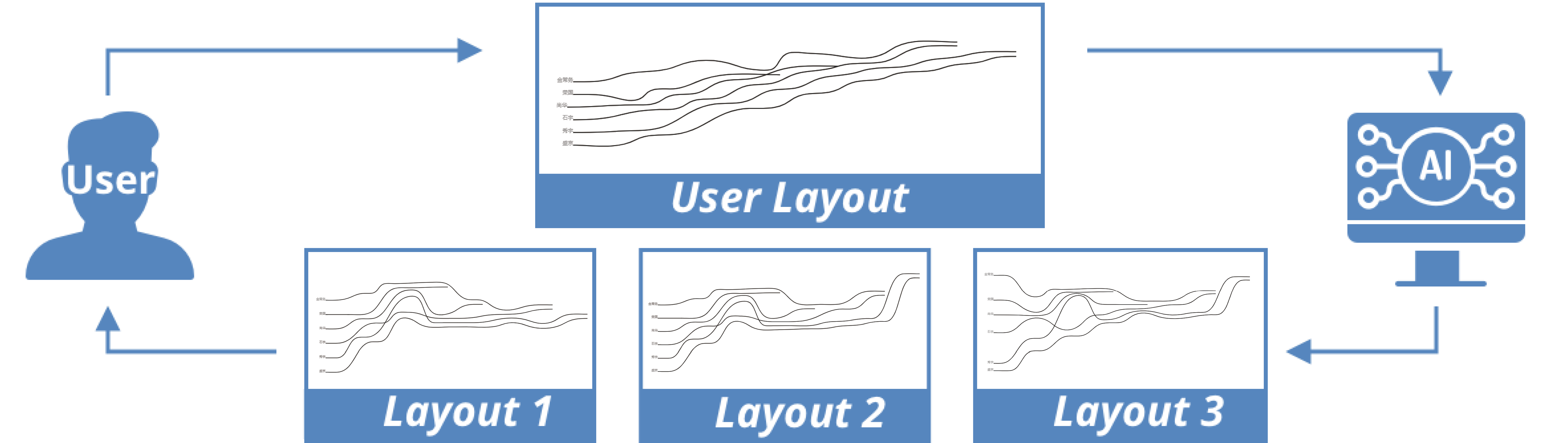}
  \caption{
    System workflow that supports a smooth and iterative co-design process between users and the AI agent.
    Users start the design process by customizing an initial storyline while the AI agent provides a set of suggestive alternative designs according to the user-specified layout.
  }
  \label{fig:workflow}
\end{figure}

\subsection{System Workflow}
Our system has two actors, namely users and AI agents (see Fig.~\ref{fig:workflow}), to support the collaborative design of storyline visualizations.
The existing storyline tools~\cite{Liu2013,Tang2018a} employ a solo-authoring workflow where users are the only actor to invoke the design process while computers mainly provide flexible design tools to ease users' efforts.
By incorporating the AI agent, we transformed the typical solo-authoring workflow into a divergent, collaborative design workflow where the agent can help users to explore the design space by providing alternative layouts.
Following \textbf{DC 1}, users should first input a story script (see Appendix A1) into the system and an initial layout would be automatically generated by the storyline optimization model~\cite{Tang2018a} which conforms to the three aesthetic goals.
Users can next modify the initial layout and then trigger the AI agent to generate various storylines proactively.
The AI-generated storylines are displayed in a list so that diverse designs can inspire users.
By default, we recommend the storyline layout which looks most similar to the user-specified one.
Next, users can simply go ahead for further refinements or smoothly switch between different AI layouts.
They can also reset to the original storyline when they are unsatisfied with the AI layouts.
Compared with the solo-authoring workflow, the co-design workflow may invoke more novel and creative ideas because both users and AI agents can contribute to the design of storylines~\cite{Yannakakis2014a}.

\begin{figure*}[ht]
  \includegraphics[width=0.95\textwidth]{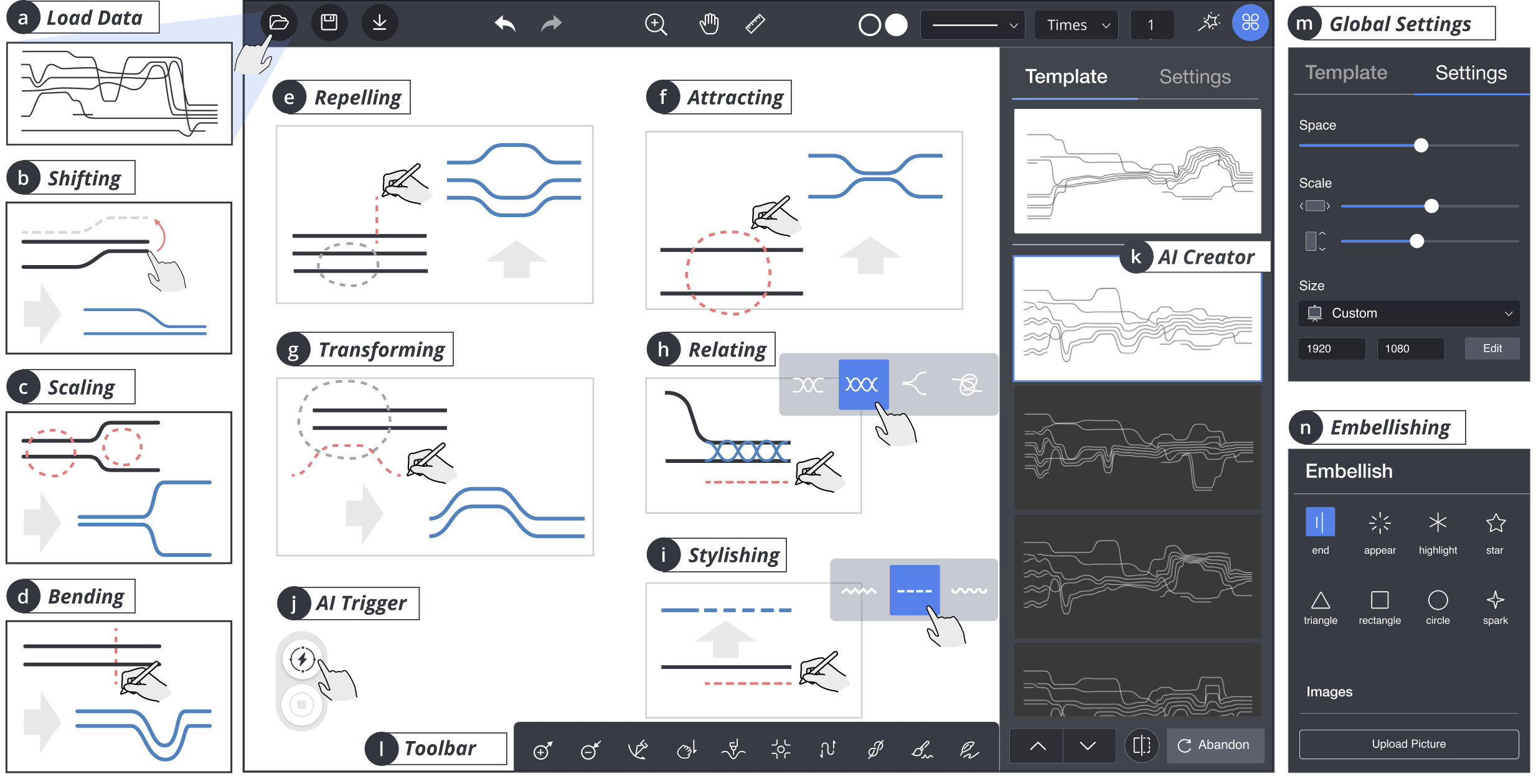}
  \centering
  \caption{
    PlotThread is composed of a menubar for (a) loading story scripts, setting canvas, and exporting storylines;
    (l) a toolbar that provides a set of easy-to-use interactions (b) to (i);
    The red lines indicate the interactions that change original layouts (black lines) into desired layouts (blue lines).
    (j) buttons for activating and stopping the AI agent and (k) a panel for presenting AI-generated layouts;
    (m) a setting panel for changing the parameters of storylines and (n) an embellishing panel for inserting icons or images into the canvas.
  }
  \label{fig:system}
\end{figure*}

\subsection{Interactions}
The core part of a mixed-initiative system entails user interactions which are vital to integrate human-AI co-creativity into the authoring process~\cite{Liapis2016a}.
To ease the difficulty of constructing storyline layouts, we first implement three high-level interactions inherited from a previous study~\cite{Tang2018a}, namely, \textit{shifting}, \textit{bending}, and \textit{scaling}.
Second, to support the design of expressive storylines, we also propose a set of novel interactions.
According to \textbf{DC2}, the new interactions should enable users to modify the overall layouts without considerable efforts in designing the individual lines.
Moreover, they should be more flexible than the high-level interactions since they do not invoke any storyline optimization model.

\subsubsection{High-level Interactions}
We only inherit the three interactions from the previous study~\cite{Tang2018a} because they can formulate user interactions as mathematical constraints which are further integrated into the optimization model to control the generation of storyline layouts.

\textbf{Shifting.}
The relationships among characters are visually revealed by the spatial proximity of the corresponding lines.
To define the characters' relationship, \textit{shifting} (Fig.~\ref{fig:system}b) enables users to drag an individual line to re-order the characters freely.

\textbf{Bending.}
The plot evolution can be indicated by the overall layout of the storyline visualizations.
For example, users can arrange the line groups in a certain direction to suggest that the story evolves into a positive or negative ending.
\textit{Bending} (Fig.~\ref{fig:system}d) enables users to easily bend a straight line into a curving line while the associated groups will be transformed automatically.

\textbf{Scaling.}
The white space can be used to present different narrative elements, such as emphasizing separations between characters to present their relationships or making room for inserted images.
\textit{Scaling} (Fig.~\ref{fig:system}c) enables users to control the size of white space between lines or groups by dragging and moving the groups of lines.

\subsubsection{Extended Interactions}
We propose four types of extended interactions to support the design space~\cite{Tang2018a} that describes the design of storylines at four levels, namely, character, relationship, plot, and structure levels. 

\textbf{Transforming} (Fig.~\ref{fig:system}g) is designed to change the overall trend of storyline layouts, \tang{which is at the plot level.}
Users should first select the scope with a circular brush, and then sketch a trajectory as the trend of the target layout.
The specified path will be segmented automatically to guide the translation of line groups of the original storyline.

\textbf{Attracting / Repelling} (Fig.~\ref{fig:system}f and ~\ref{fig:system}e) are designed to indicate the closeness between the line groups, \tang{which is at the structure level.}
After selecting line groups with a circular brush, users can draw a straight line to indicate whether the selected lines should be attracted or repelled.

\textbf{Relating} (Fig.~\ref{fig:system}h) is designed to assist users in visually presenting the relationships among characters using various visual elements, such as merged or twined lines~\cite{Tang2018a}, \tang{which is at the relationship level.}
After selecting the desired visual elements, users can choose the group of lines they want to relate with each other.

\textbf{Stylishing} (Fig.~\ref{fig:system}i) is developed for users to decorate the lines with diverse stroke styles (e.g., dash and zigzag), \tang{which is at the character level}.
Users first need to select a line style and then brush the target line which they want to embellish.
Users can also highlight certain events or characters by directly inserting graphics or icons.

\subsection{Interface}
Users first need to load data (Fig.~\ref{fig:system}a) that are scripts recording characters and their scenic interactions.
The AI-based creator can be triggered (Fig.~\ref{fig:system}j) at any time during the authoring process and then provides a list of suggestive layouts based on users' layouts.
As shown in Fig.~\ref{fig:system}k, the user-specified layout is shown at the top of the list to be compared with alternative layout designs.
To inspire lateral thinking~\cite{De2010a}, we not only present the ``final'' layout that looks most similar to the user layout but also exhibit the intermediate layouts that demonstrate how the AI agent modifies storylines.
Users can freely browse and adopt alternative layouts.
We develop two panels (Fig.~\ref{fig:system}m and ~\ref{fig:system}n) to support the creative design of storylines where users can \textit{insert} images, \textit{add} icons, and \textit{change} model parameters.

\section{Problem Overview}  
Following the two principles (\textbf{D1} and \textbf{D2}) of collaborative design, our goal is to train an agent that learns how to resemble users' intermediate layouts using a set of user-shared interactions.
Furthermore, we want to leverage the aesthetic goals (\textbf{G1} to \textbf{G3}) to produce well-optimized layouts.
Thus, the agent is trained to predict the high-level interactions that can modify an automatically-optimized layout to resemble a user-created layout.
The high-level interactions preserve the aesthetic quality of layouts as much as possible since we employ the optimization model~\cite{Liu2013} to re-generate storyline layouts.

The problem is formulated as follows:
given a user layout $L_u$ crafted by a user and an origin layout $L_o$ generated by the optimization model~\cite{Liu2013}, the agent predicts the actions used to modify $L_o$ according to $L_u$.
Like human designers, the agent is designed to predict the ``next'' action by observing the ``current'' layout and imitating the user layout.
To avoid local minima, the agent should balance current actions and future actions by maximizing the cumulative rewards after finishing the given number of actions, rather than the current gain.
Inspired by the similar task of reproducing paintings~\cite{Huang2019a}, we employ reinforcement learning to achieve this long-term delayed-reward design.

\section{Reinforcement Learning}
In this section, we describe the entire process for designing the reinforcement learning framework from constructing storyline layouts, generating training datasets, building neural networks, and learning the storyline agent.

\subsection{Storyline Layout Construction}
Storyline visualizations depict how characters interact with each other.
Generally, each line represents a character, and a group of lines indicates that the associated characters are together at a given time slot~\cite{xkcd}.
Given a story with $N$ characters and $M$ time slots, the path of the $i$-th character $C_i$ can be described as a sequence of points $[y_i^{0},y_i^{1}...,y_i^{M-1}]$.
The overall layout can be denoted as a set of characters $L=\{C_i\}_{i=0}^{N-1}$ which can be further depicted as

\begin{equation}
  M_{pos}=[y_{i}^{j}]_{i=0,...,N-1;j=0,...,M-1}
  \label{eq:matPos}
\end{equation}

The main difficulty in obtaining a storyline layout is the calculation of its position matrix $M_{pos}$, which pursues the maximization of the aesthetic metrics (\textbf{G1} to \textbf{G3}) while satisfying the primary narrative constraints (\textbf{C1} and \textbf{C2}).
Given that the performance of the existing storyline algorithms~\cite{Liu2013,Tanahashi2015,Tang2018a} has been considerably improved, we adopt iStoryline~\cite{Tang2018a} as the renderer to calculate the layout.
First, iStoryline is implemented on the basis of StoryFlow to achieve a real-time generation for a large collection of storylines~\cite{Liu2013}, which is vital for training an agent that needs to reproduce storylines for over thousands of hundreds of times.
Second, iStoryline extends the optimization model of StoryFlow to integrate a more diverse set of narrative constraints, which is crucial for the agent to fully explore the overall design space and customize storyline layouts without losing too much aesthetic quality.

In PlotThread, we inherit three high-level interactions, namely, \textit{shifting}, \textit{bending}, \textit{scaling}, from iStoryline, which insert three novel types of narrative constraints to the three optimization stages, namely \textit{ordering}, \textit{alignment}, and \textit{compaction}~\cite{Tang2018a}.
Next, we introduce how these high-level interactions are incorporated into the optimization model to enable an efficient customization of storyline layouts.

\textbf{Shifting} determines the vertical order of characters using a constrained crossing minimization algorithm~\cite{Forster2004}, which generates \textit{ordering} constraints using a set of order pairs $[o_i^j,o_{i^{'}}^{j}]_{i,i^{'}< N;j< M}$ where $o_i^j$ indicates the order of the $i$-th character at the $j$-th time slot.
The constraint suggests that the $i$-th character should be ``ahead'' of the $i^{'}$-th character at the $j$-th time slot.
After solving the ordering algorithm~\cite{Forster2004}, the order of characters during the whole timeline can be obtained using

\begin{equation}
  M_{order}=[o_{i}^{j}]_{i=0,...,N-1;j=0,...,M-1}
  \label{eq:matOrder}
\end{equation}

\textbf{Bending} determines the straightness of characters along the timeline via the dynamic programming algorithm~\cite{Liu2013}, which generates \textit{alignment} constraints using a set of indicators $[e_i^j]_{i< N;j< M}$.
The variable $e_{i}^{j}$ is set to $1$ when the $i$-th character are aligned at both the $j$-th and its previous time slots.
By default, the indicators at the first time slot are set to $1$ so that $\{e_i^0=1\}_{i=0,...,N-1}$.
After solving the dynamic programming~\cite{Liu2013}, the alignment situations can be obtained using

\begin{equation}
  M_{align}=[e_{i}^{j}]_{i=0,...,N-1;j=0,...,M-1}
  \label{eq:matAlign}
\end{equation}

\textbf{Scaling} determines the white space among characters through the least-square method~\cite{Liu2013}, which generates \textit{compaction} constraints using a set of inequalities $\{d_{1}<|y_{i}^{j}-y_{i-1}^{j}|<d_{2}\}_{i< N;j< M}$ where $d_1$ and $d_2$ are numerical values to indicate the lower and upper bounds of the white space among the $i$-th and its last characters.
After obtaining the results (Eq.~\ref{eq:matOrder} and~\ref{eq:matAlign}) of the two previous optimization stages, the position matrix (Eq.~\ref{eq:matPos}) can be obtained by solving a constrained convex optimization problem which is detailed \tang{in Appendix A2}.

\subsection{Training Data Collection}
Training neural networks require a large number of high-quality datasets~\cite{Kwon2019a,liu2018steering}.
Although Tang et al.~\cite{Tang2018a} have extended the collection of hand-drawn storyline illustrations, the size of the dataset is too small for a machine learning task.
Moreover, the manual production of training data is a labor-intensive task which requires considerate time and human resource.
Inspired by the recent studies on graph drawings~\cite{Wang2019a,Kwon2019a}, we generate a set of well-optimized storyline layouts using the optimization model~\cite{Liu2013}.
Although automatically-generated storylines are not comparable to the hand-drawn illustrations in terms of both aesthetic quality and expressiveness~\cite{Tang2018a}, our goal is to train an agent that can imitate users’ layouts instead of generating storylines that are comparable or superior to hand-drawn ones.

To obtain considerate and diverse datasets, researchers have employed a grid search that applies different combinations of random parameters on the graph models~\cite{Kwon2019a,Wang2019a}.
Following this common practice, we use iStoryline~\cite{Tang2018a} to generate the training datasets due to its ability to produce aesthetic storyline layouts in a short time.
Notice that iStoryline only receives two parameters, namely \textit{inner gap} and \textit{outer gap}, to determine the white space between individual lines and the groups of lines, respectively.
Thus, merely modifying the model parameters cannot produce sufficient storylines with diverse layouts.
We apply random searching in generating different narrative constraints described in Sec. 5.1, which are further integrated into the optimization model~\cite{Liu2013} to control the diversity of storyline layouts.
Mathematically, the training data is a set of storyline pairs $<L_o,L_u>$ where $L_o$ is the origin layout generated by the optimization model directly, and $L_u$ is the ``user'' layout simulated by inserting randomly-selected narrative constraints into the optimization model~\cite{Liu2013}.
However, the simulated ``user'' layout $L_u$ may not be visually ``better'' than the origin layout $L_o$ because more narrative constraints are used to restrict the optimization of storylines.
Since the goal of the AI agent is to provide a list of possible layouts according to users' layouts, the key of our RL model is to teach the agent to refine origin layouts and imitate users' layouts instead of producing extremely-optimized storylines.

Following the design considerations mentioned above, we first extract story scripts that describe characters and their scenic interactions from the hand-drawn illustrations\footnotemark[1].
Each story script records a set of time slots that indicate who are together at a given time.
To ensure the diversity of training data, we evenly produce three groups of narrative constraints, namely, ordering, alignment, and compaction constraints with random parameters.
We then randomly select $K$ constraints from the three groups to obtain different layouts for the same story script.
The selected constraints are the ground truth that the agent needs to learn and predict when modifying origin layouts $L_o$ to imitate user layouts $L_u$.
The variable $K$ indicates how many steps the agent can have to reproduce user layouts.
In our case, we set $K=15$ because the agent should complete the authoring task within reasonable time to avoid losing users' attention.
We obtain $20$ story scripts from the published gallery and generate $1000$ storyline layouts for each story.
In total, we generate $20000$ layouts to train the AI agent.

\footnotetext[1]{\url{https://istoryline.github.io/gallery/}}

\subsection{Model Architecture}
Given a user layout $L_u$ and an origin layout $L_o$, the agent aims to predict a sequence of actions $\{a_k\}_{k=0}^{K-1}$ where rendering $a_k$ on $L^{(k)}$ leads to $L^{(k+1)}$.
The initial layout $L^{(0)}$ can be obtained from the origin layout $L_o$.
The final layout $L^{(K-1)}$ can be obtained by rendering the consecutive actions, which should be visually similar to $L_u$ as much as possible.
This design issue can be formulated as a \textit{Markov Decision Process}~\cite{Huang2019a} with a state space $\mathbb{S}$, an action space $\mathbb{A}$, a transition function $T(s_t,a_u)$ and a reward function $R(s_t,a_u)$~\cite{Racaniere2017a}.

\subsubsection{State and Transition Function}
The state space describes all possible layouts that an agent can obtain after rendering actions.
Mathematically, we define a state $s_u=(L^{(k)},L_u,k)$ where $L^{(k)}$ and $L_u$ refer to the layouts that can be represented by the position matrix $M_{pos}$ and the variable $k$ indicates the $k$-th step for the agent.
We further define the transition function $s_{k+1}=T(s_k,a_k)$ that describes the transition process between states $s_k$ and $s_{k+1}$, which is implemented by applying action $a_k$ on state $s_k$.

\subsubsection{Action}
To support the collaborative design, we define the action space as the high-level interactions (discussed in Sec. 5.1) for three reasons.
First, it is necessary for the agent to share the same action space with users so that they can work concurrently to design storyline visualizations.
Second, the high-level interactions are implemented on the basis of the constrained optimization model~\cite{Liu2013,Tang2018a} so that the agent can produce well-optimized layouts in terms of the aesthetic goals.
Third, it is sufficient to modify storyline layouts with these interactions so that we do not include the other interactions proposed in PlotThread.
Formally, an action $a_k$ of the storyline agent is a set of parameters that define a narrative constraint (e.g., ordering, alignment, compaction constraint).
The behaviors of the agent can be further described using a policy function $P: \mathbb{S}\rightarrow\mathbb{A}$ that maps states to deterministic actions\cite{Sutton2018a}.
After predicting action $a_k$ at step $k$, the state can evolve using the transition function $s_{k+1}=T(a_k,s_k)$, which runs for $K$ steps~\cite{Yang2017a}.

\begin{figure}
  \includegraphics[width=0.5\textwidth]{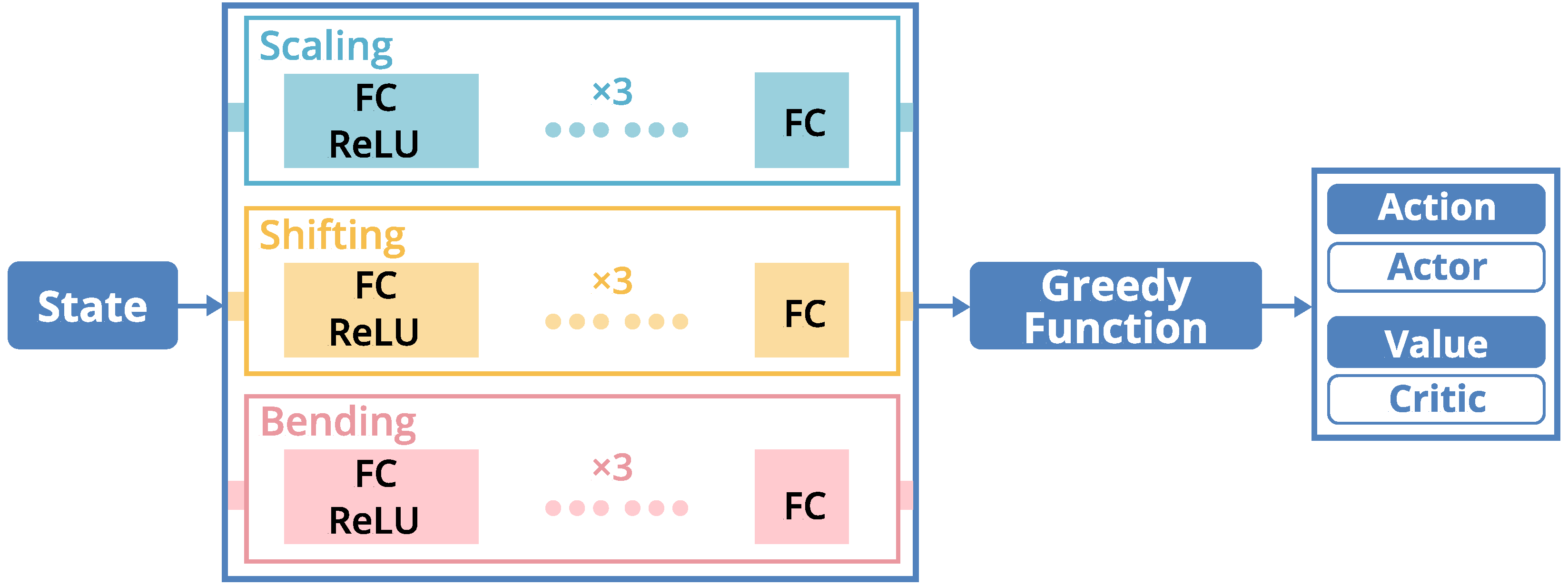}
  \caption{
    Neural network architecture for the AI agent.
    \textbf{FC} refers to fully-connected layer, and \textbf{ReLU} represents an activation function.
    Three neural networks are employed to separately predict the three high-level interactions.
    A greedy function is used to obtain final actions and values.
  }
  \label{fig:network}
  \vspace{-5mm}
\end{figure}

\subsubsection{Reward}
Reward is a stimulus for the agent to improve its prediction ability~\cite{Mnih2015a}.
Our goal is to guide the agent to resemble the layout on which users are working and produce alternative layouts to inspire co-creativity.
We formulate the reward as the similarity between the user layout $L_u$ and the layout $L^{(k)}$ produced by the agent at step $k$.
To quantify the layout similarity, we follow the well-established framework~\cite{Liu2013} that measures storyline layouts in three aspects:

\textbf{Ordering Feature}
The first step to obtain an aesthetic storyline layout is to determine the vertical order of characters.
The \textit{ordering} variable $o_{i}^{j}(L)$ indicates the ranking position of the $i$-th character at the $j$-th time slot for the layout $L$.
Based on that, we formulate the ordering similarity between the user layout $L_u$ and the layout $L^{(k)}$ at step $k$ as

\begin{equation}
  S_{order}^{(k)} = Comp(M_{order}^{L_u}, M_{order}^{L^{(k)}})
  \label{eq:orderDiff}
\end{equation}

\textbf{Alignment Feature}
After obtaining the orders of characters, the second step is to determine the alignment situation of characters along the whole timeline.
Given a layout $L$, the \textit{alignment} variable $e_i^j(L)$ indicates whether the $i$-th character is aligned at the $j$-th time slot and the previous slot.
Following the same mathematical notations, we quantify the alignment similarity as
\begin{equation}
  S_{align}^{(k)} = Comp(M_{align}^{L_u}, M_{align}^{L^{(k)}})
  \label{eq:alignDiff}
\end{equation}

\textbf{Position Feature}
The last step for generating storyline layouts is to calculate the exact positions of characters by minimizing the white space of the overall layout.
The \textit{position} variable $e_i^j(L)$ suggests that the position of the $i$-th character at the $j$-th time slot in the layout $L$.
We calculate the position difference of the two layouts using

\begin{equation}
  D_{pos}^{(k)} = Dist(M_{pos}^{L_u}, M_{pos}^{L^{(k)}})
  \label{eq:positionDiff}
\end{equation}

where $Comp(\cdot)$ is a counting function that self increment one if the corresponding values of two matrices are the same and $Dist(\cdot)$ is a distance function that calculates the difference between two matrices using Euclidean metric.
We further employ sigmoid function $\overline{S}(\cdot)$ to normalize the three visual features.
Based on that, we define the similarity between the user layout $L_u$ and the $k$-th step layout $L^{(k)}$ using a linear scheme $S(k)=\omega_1\overline{S}(S_{order}^{(k)})+\omega_2\overline{S}(S_{align}^{(k)})+\omega_3\overline{S}(D_{pos}^{(k)})$.
The reward at $k$-th step can be obtained using $r(s_k,a_k)=S(k)-S(k+1)$.
To make the final result resemble the user layout, we maximize the cumulative rewards in the whole episode using a discounted scheme that $R_k=\sum_{k^{'}=k}^{K}\gamma^{k^{'}} r(s_{k^{'}},a_{k^{'}})$ with a discounting factor $\gamma\in [0,1]$.
The default parameters $[\omega_1,\omega_2,\omega_3,\gamma]$ are set to $1$.

\subsubsection{Network Architecture}
Due to the high variability and complexity of narratives, we first normalize the input layout into a $H\times H$ matrix which can be regarded as a one-channel image (By default, we set $H=100$).
To extract the visual features from storyline layouts, we employ the network structure that is similar to ResNet-18~\cite{He2016a}.
Given that storyline layouts are less complicated than real-world images, we simplify the network structure by removing all convolution layers to preserve visual information.
In our experiments, we discover that the fully-connected layers are capable of predicting actions for generating storyline layouts.
To ease the difficulty of exploring the mix-type action space and stabilize learning, we separate the network architecture into three parallel components~\cite{Kempka2016a} that aim at exploring the different parts of the action space.
Specifically, every component is designed only to explore the action space of one of the high-level interactions (see Fig.~\ref{fig:network}).
\tang{In the end, we employ a greedy function to calculate the reward and determine the action.}

\begin{equation}
  \tang{R_k=\sum_{k^{'}=k}^{K}\gamma^{k^{'}} \max_{a_{k^{'}}} r(s_{k^{'}},a_{k^{'}})}
  \label{eq:greedy}
\end{equation}

\begin{figure}
  \includegraphics[width=0.5\textwidth]{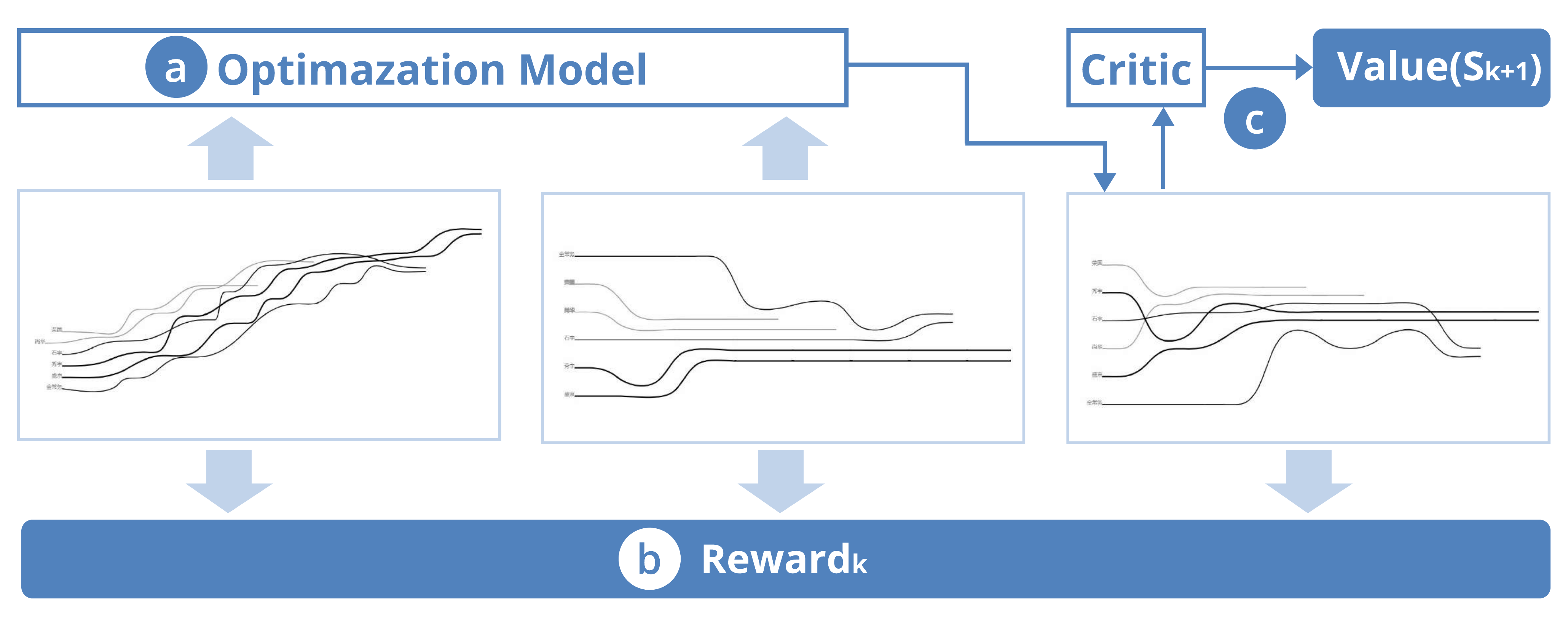}
  \caption{Learning algorithm for the AI agent:
  (a) use the optimization model~\cite{Liu2013} to produce storylines;
  (b) measure the similarity between user-specified and ``current'' layouts to obtain the reward;
  (c) calculate the critic value to predict the ``next'' action.}
  \label{fig:learning}
  \vspace{-5mm}
\end{figure}

\begin{figure*}[ht]
  \includegraphics[width=0.95\textwidth]{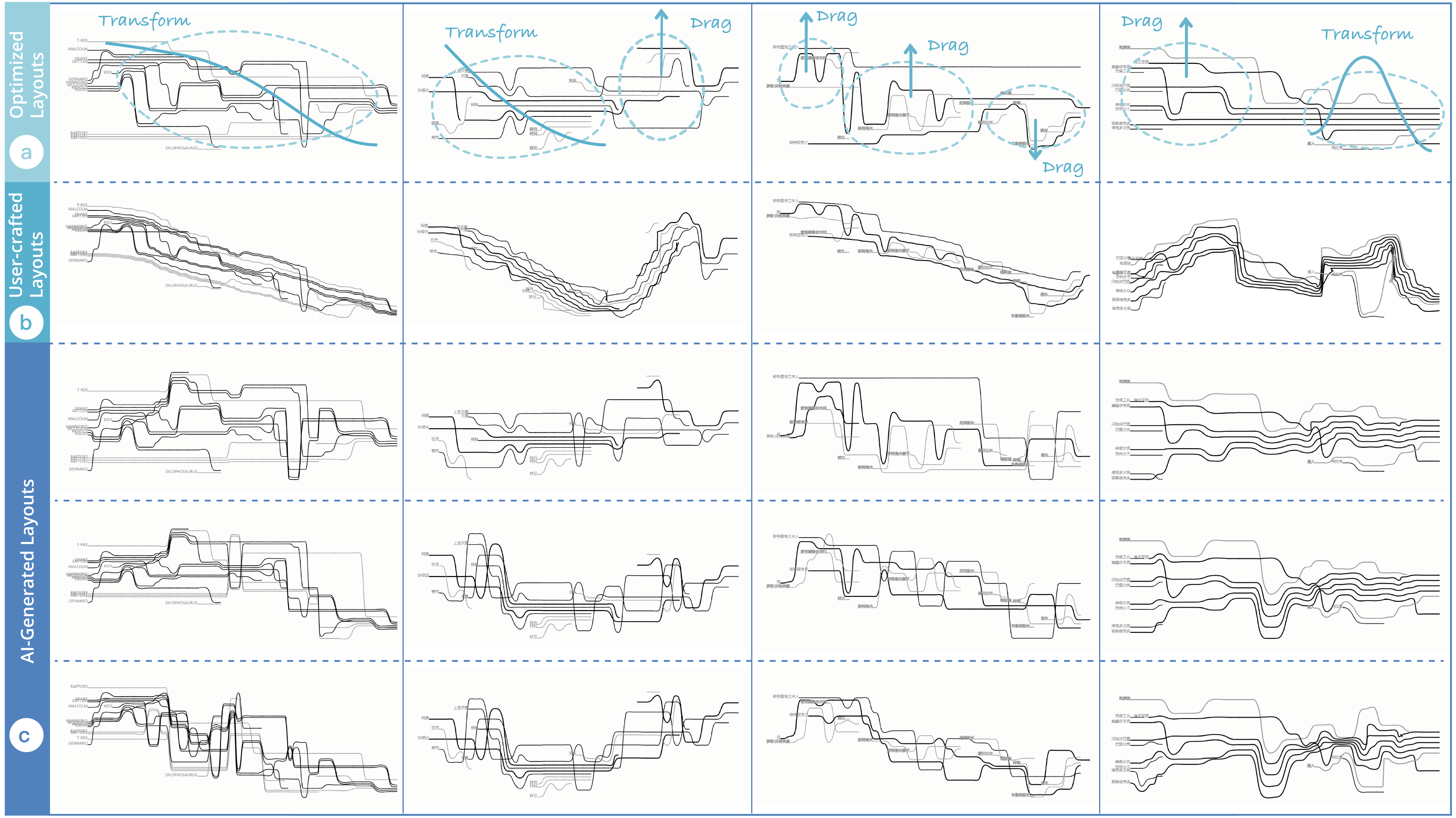}
  \centering
  \caption{Qualitative experiments: (a) the initially-optimized layouts generated by the optimization model~\cite{Liu2013}.
  (b) the user layouts modified by the interactions shown in (a).
  (c) the AI-generated layouts that resemble the user layouts but with improved aesthetic quality.
  The intermediate layouts at the $k$-th step ($k=1,5,15$) are also presented to indicate how the AI agent reproduces the user layouts.
  The four cases are \textit{Jurassic Park}, \textit{WuKong}, \textit{Moon and Sixpence}, \textit{Justice League} (from left to right).
  }
  \label{fig:model}
  \vspace{-5mm}
\end{figure*}

\subsection{Learning}
We first introduce the standard setting for reinforcement learning~\cite{Sutton2000a} where an agent interacts with an environment over a certain number of time steps,
and then describe how to train the agent using the state-of-the-art framework, namely, asynchronous advantage actor-critic~\cite{Mnih2016a}.

In a typical actor-critic model~\cite{Peters2008a}, researchers usually employ two neural networks to present \textit{actor} and \textit{critic}, respectively (see Fig.~\ref{fig:learning}).
An actor observes the environment by receiving an state $s_k$ and then predict an action $a_k$ at time step $k$, while a critic obtains the state $s_k$ to predict cumulative reward in the future.
In general, the policy function $\pi(a_t|s_t,\theta_{\pi})$ characterizes the actor's behaviors which can be formulated as a mathematical probability function.
Since an agent aims to maximize the expected cumulative reward~\cite{Kaelbling1996a}, the value of state $s_k$ under policy $\pi$ can be defined as $V^{\pi}(s,\theta_{V})=\mathbb{E}(R_k|s_k=s)$ that is the expected return for following policy $\pi$ from state $s$.
Hence, the problem of training a storyline agent is to obtain the parameters $(\theta_{\pi},\theta_{V})$ of the neural networks for the policy function $\pi$ and the value function $V$.
The updates on the model parameters~\cite{Bhatnagar2009a} can be written as

\begin{align*}
  \Delta\theta_{\pi} & \leftarrow \Delta\theta_{\pi} + \nabla_{\theta_{\pi}}log\pi(a_k|s_k; \theta_{\pi}) (R_k-V(s_k;\theta_{V})) \\
  \Delta\theta_{V} & \leftarrow \Delta\theta_{V} + \frac{\partial(R_k-V(s_k; \theta_{V}))^2}{\partial \theta_{V}}
\end{align*}

where $\Delta\theta_{\pi}$ and $\Delta\theta_{V}$ are the updates applied to the model parameter $\theta_{\pi}$ and $\theta_{V}$, respectively.

However, training an agent in a complicated high-dimensional mix-type action space is difficult due to the unstable learning problem and the requirements of large computational resources~\cite{Mnih2016a,Zhou2019Abductive}.
To overcome these issues, Minih et al.~\cite{Mnih2016a} propose a novel asynchronous framework that enhances the existing RL algorithms, such as Q-learning~\cite{Lillicrap2016a}, and actor-critic methods~\cite{Peters2008a}.
The key idea is to use asynchronous actor-learners that run in parallel to explore different parts of the environment~\cite{Mnih2016a,liu2017towards}.
Instead of using different machines, the actor-learners are running on the different processes to remove the communication costs and improve training efficiency.
Moreover, the researchers observe that it is more likely for the multiple actor-learners to be uncorrelated than a single agent when applying the overall changes to the model parameters.
The updates applied to the parallel agent~\cite{Mnih2016a} will be updated on the main agent to combine the asynchronous changes of the model parameters on different processes.

\begin{figure*}[ht]
  \includegraphics[width=0.95\textwidth]{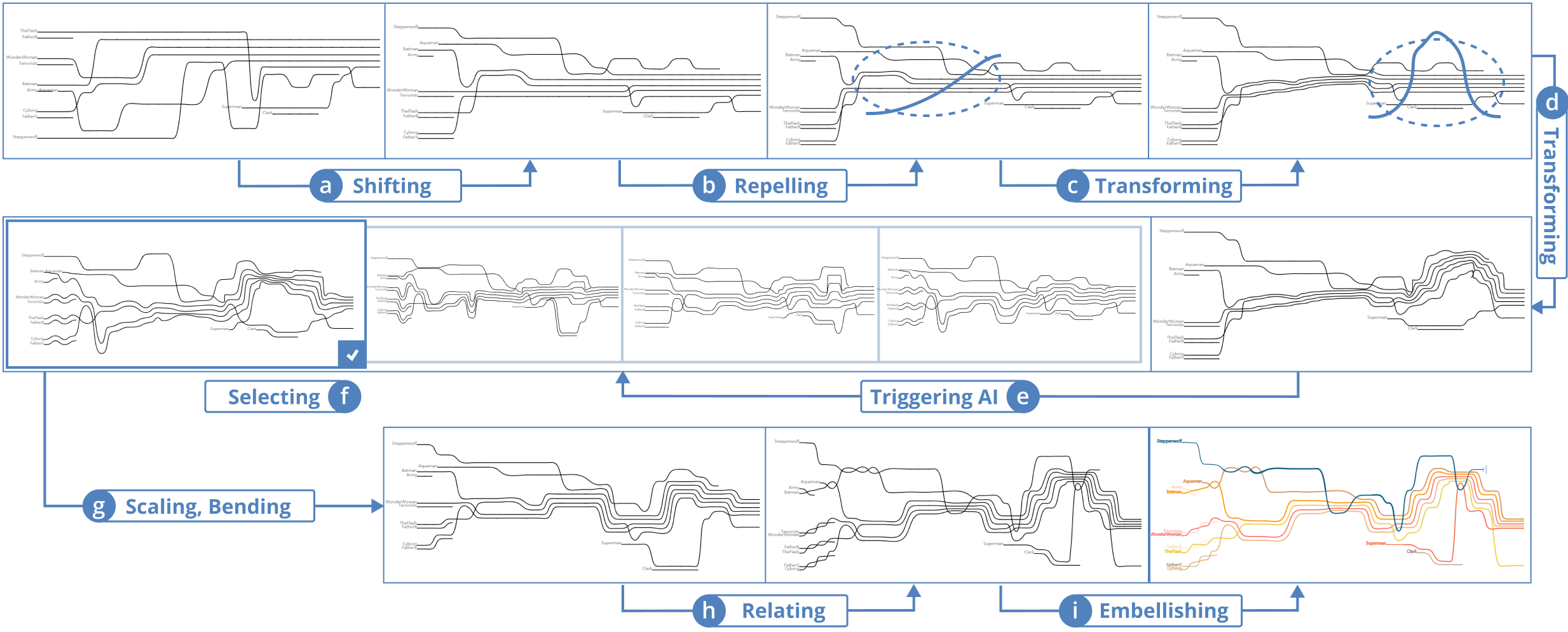}
  \centering
  \caption{
    This case illustrates the authoring process of the storyline visualization (\textit{Justice League}) using PlotThread.
    The designer first customizes an initial layout through (a) \textit{Shfiting}, (b) \textit{Repelling}, (c) and (d) \textit{Transforming} (\tang{Dashed ellipses indicate the transforming regions and solid paths represent the transforming shapes}).
    The AI agent is then (e) triggered to produce suggestive storylines and (f) the desired one is selected.
    S/He further improves the AI-generated layout using high-level interactions, namely \textit{Scaling} and \textit{Bending} (g).
    The relationships among characters are revealed using (h) \textit{Relating}, and the layout is (i) embellished to enrich the narration.
  }
  \label{fig:justice}
  \vspace{-4mm}
\end{figure*}

\section{Experiments}
\textbf{Implementation.}
We employ a client-server architecture to develop PlotThread.
The web interface is implemented using TypeScript~\cite{TypeScript} and ECharts.js~\cite{ECharts} while the server side is implemented using Python and the popular machine learning library PyTorch~\cite{PyTorch}.
To support the flexible customization of storyline visualizations, we adopt the well-established graphic library, namely Paper.js~\cite{Paperjs}.
We also develop a storyline layout optimizer which is implemented using C\# to modularize PlotThread.

To validate the effectiveness of the reinforcement learning (\textbf{RL}) model, we conduct both quantitative and qualitative experiments on four datasets.
The input stories are visualized using the optimization model~\cite{Liu2013} in Fig.~\ref{fig:model}a.
We first show that the agent has leveraged both aesthetic and expressiveness in producing various types of storyline layouts.
To simulate the real authoring process, we create four user layouts, including \textit{incline-}, \textit{stair-}, and \textit{wave-layout} (see Fig.~\ref{fig:model}b), using the extended interactions that reshape the overall layouts without invoking optimization models.
Apparently, the user layouts are twisted and do not satisfy the aesthetic goals, but they are regarded as more expressive in terms of the diverse visual forms.
We then invoke the RL model to predict actions that can modify the initially-optimized layouts (Fig.~\ref{fig:model}a) to resemble the user layouts (Fig.~\ref{fig:model}b).
The results (Fig.~\ref{fig:model}c) indicate that our RL model can successfully capture the visual features from the user layouts and produce more expressive layouts than the initially-optimized ones.
Despite that the AI-generated layouts seem to have more edge crossings than the initially-optimized layouts, they still preserve a satisfactory aesthetic quality compared to the user layouts.
Thus, we believe our agent achieves a better trade-off between expressiveness and aesthetics even though it increases expressiveness at the cost of some aesthetic quality.

\begin{figure}[ht]
  \includegraphics[width=0.48\textwidth]{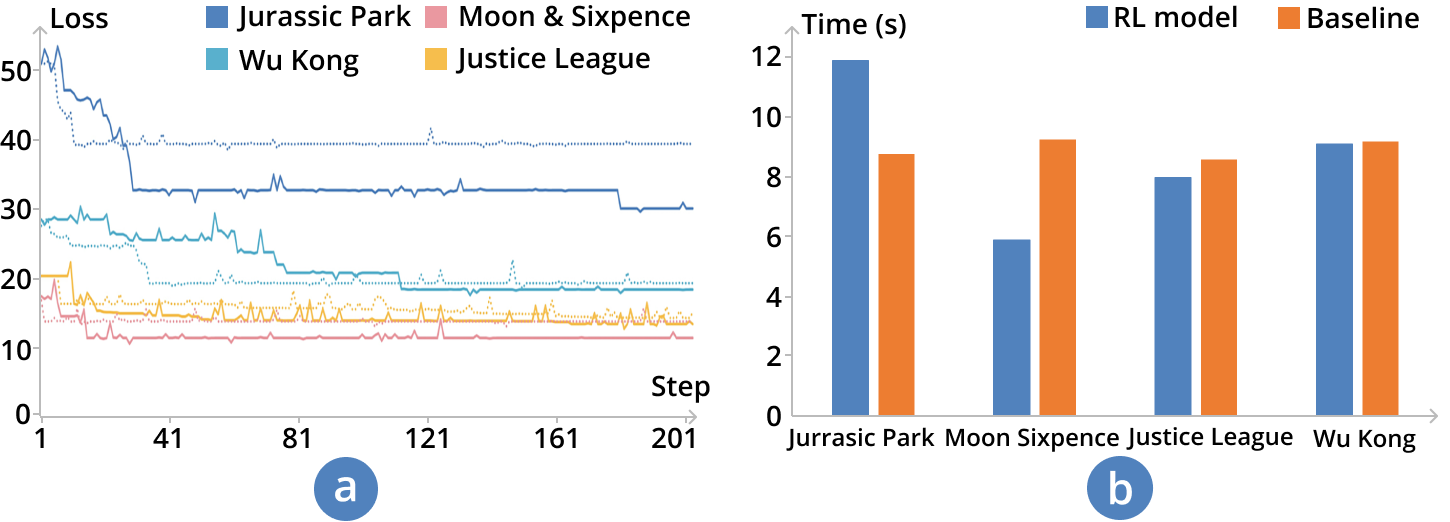}
  \caption{
    \tang{
      Quantitative experiments: (a) x-axis represents the number of steps, and y-axis indicates the loss that is inverse of the cumulative rewards.
      The solid and dashed lines indicate the performance of the RL model and the baseline method, respectively.
      (b) y-axis indicates the running time of the RL model (blue) and the baseline method (Orange).
    }
  }
  \label{fig:experiment-quantitative}
  \vspace{-4mm}
\end{figure}

We also conducted quantitative experiments on a desktop with a CPU (3.7GHz) to evaluate the search power and \tang{time performance} of the AI agent by comparing it with a baseline method.
We repeated the experiments $4$ times and calculated average values to avoid the influences of CPU scheduling.  
Since there are no prior RL models on designing storyline visualizations, we implemented a greedy algorithm that randomly selects a group of actions and then adopts the one that can improve the reward.
We compare the AI agent with the greedy algorithm by measuring their convergence rates and \tang{time} when performing the same tasks.
As shown in Fig.~\ref{fig:experiment-quantitative}a, the baseline method can improve the rewards dramatically in the short term but they are trapped in the local optimums finally.
While the AI agent seems to have difficulties in searching the design space in the beginning, it finally achieves better performances than the baseline method in the long term.
The results indicate that our RL model has successfully learned how to ``think'' when designing storylines and can sacrifice short-term rewards to achieve long-term planning.
Moreover, both the AI agent and the baseline method can converge to final layouts within $12$ seconds (see Fig.~\ref{fig:experiment-quantitative}b).
In PlotThread, we set the default steps of the agent to $15$ which ensures a satisfactory response time for users' interactions.

\section{Use Cases}
In this section, we illustrate the usage of PlotThread.
A storyline visualization called \textit{Justice League} is created to describe the authoring process of customizing the layout with the assistance of the AI agent.
This use case indicates how people and the agent can work together to achieves users' design requirements.
Following the same procedure, we create more use cases (see Fig.~\ref{fig:teaser}) to demonstrate that PlotThread can be used to design various stories and produce diverse layouts.

As a proof of concept, we simplify the story of Justice League~\cite{justice} and only depict events that are vital for the evolution of the narrative.
The story depicts how superheroes stand together and establish Justice League to fight against Steppenwolf, which can be roughly divided into three stages.
First, Batman and Wonder Woman decided to recruit team members.
They recruited The Flash, Cyborg, and Aquaman to save the world from the catastrophic threat caused by Steppenwolf.
The second stage begins with the fight against Steppenwolf's invasion and the rescue act for Superman.
But Superman attacked the other superheroes who try to rescue him since Steppenwolf had twisted his mind.
After he recovered, he decided to join Justice League.
The third stage was the climax where the superheroes struggled with the fight against Steppenwolf, and finally won with the return of Superman.


As shown in Fig.~\ref{fig:justice}, we illustrate how to create the storyline visualization step by step using PlotThread.
We use ``George'' who refers to the user to describe the authoring process.
George first loads the story data which is a formatted document for the storyline renderer~\cite{Tang2018a}, and he obtains an initial storyline.
Then, he uses \textit{Shifting} to change the order of characters, and separate Steppenwolf and the members of Justice League (Fig.~\ref{fig:justice}a).
Based on his understanding of the story, he wants to transform the layout of the first stage into an ``up-step'' shape, where Batman and Wonder Woman tried to increase their team by recruiting new superheroes.
He uses \textit{Transforming}, selects the superheroes involved, and draws an ascending step-like line to obtain the initial step layout (Fig.~\ref{fig:justice}c).
Next, he moves on to transform the shape of other stages, for example, in the climax stage, he draws a parabolic curve (Fig.~\ref{fig:justice}d) to illustrate how Justice League beats Steppenwolf.
George thinks the appearance is not legible or aesthetically appealing enough, so he seeks for AI assistance by triggering the AI creator.
After AI gives the results, he quickly switches between different layouts and finds one (Fig.~\ref{fig:justice}f) which has few crossings and deviations but preserving the narrative trends specified by him.

George continues to clarify the relationship between the groups of characters in a more coarse-grained way.
Using \textit{Repelling}, he emphasizes the part when Superman leaves the other superheroes, and when he reunites (Fig.~\ref{fig:justice}b).
He triggers the AI creator when he wants to improve the appearance and gain some inspirations about the storyline design (Fig.~\ref{fig:justice}e).
After clarifying the trend of the story as well as the general narrative structure, George starts to work on detailed refinements using \textit{Bending} and \textit{Scaling} (Fig.~\ref{fig:justice}g).
For example, he emphasizes the closeness of Justice League at the end of the whole story using \textit{Scaling}.
After completing the layout, George uses \textit{Relating} to embellish the plots and make them more expressive (Fig.~\ref{fig:justice}h).
For example, he uses twined lines to illustrate the intense fights.
Finally, he embellishes the picture by adding icons, changing line colors and stroke weight, as well as adding text annotations (Fig.~\ref{fig:justice}i).

\section{User Feedback}
To evaluate the effectiveness and usability of PlotThread, we conducted semi-structured interviews with three experts.
The first expert (\textbf{EA}) was an artist who graduated from a national academy of art.
She evaluated the output storylines crafted by PlotThread (see Fig.~\ref{fig:teaser}) and compared them with the storylines generated by the optimization model~\cite{Liu2013} and human artists~\cite{Tang2018a}.
The second expert (\textbf{ED}) was a professional UI/UX designer who worked for an international software company.
She helped to test the usability of PlotThread because she had rich experiences in using various commercial design tools (e.g., Adobe AI/PS).
The third expert (\textbf{EV}) was a senior researcher who studied visualization and visual analytics~\cite{Munzner2014a} for eight years.
He evaluated the system development of PlotThread and discussed the potential applications for storyline visualizations.
The interview includes a 30-min demonstration of PlotThread and a 30-min discussion.

\textbf{EA} mainly evaluated the storylines generated by different agents from the aesthetic and narrative aspects.
We provided three kinds of storylines: the AI-assisted storylines created by PlotThread (see Fig.~\ref{fig:teaser}), the extremely-optimized storylines generated by the optimization model without human involvement~\cite{Liu2013}, and the hand-drawn storylines created by artists~\cite{Tang2018a}.
She thoroughly compared the visual designs of the different storylines and surprisingly found that the extremely-optimized storylines are hardest to read although they have fewest crossings and deviations.
She inferred that ``viewers intend to pay attention to line groups which are hard to be distinguished in extremely-optimized storylines because they are too compact.''
This observation validates the effectiveness of PlotThread which intends to better balance the aesthetic goals and the narrative constraints by sacrificing some aesthetic quality to enrich the narration.
On the one hand, the AI agent is inherently driven by the optimization model so that it can produce well-optimized results.
On the other hand, the agent resembles the input layouts which can be flexibly customized to indicate more narrative details.

\textbf{ED} mainly focused on the system design, including the human-AI collaborative workflow, the design of interactions, as well as the user interface.
She was impressed by the interaction and interface design and commented that ``the interactions are intuitive and the interface is easy to follow,'' but she also pointed out that users may need some training when they first use the system.
To lower the learning cost, we will further improve the system with a user-friendly built-in user guide.
When asked about the experience of human-AI collaboration, she commented that ``users may doubt whether the AI agent can really understand their intentions, so they may be very reluctant to seek AI for help.''
This concern reveals a common trust issue widely existing in ``black box'' models.
One possible solution is to provide an animation that demonstrates the evolution of layouts and how the AI agent modifies storylines.
She also suggested that it would be helpful if users do not need to prepare story scripts because ``it will challenge general users who do not have story scripts.''
Additionally, she provided a potential application of PlotThread that ``it may be promising for preschool teachers to tell stories visually using PlotThread.''

\textbf{EV} commented on the performance of the reinforcement learning algorithm and the applicability of PlotThread.
He confirmed the effectiveness and expressiveness of the storylines (see Fig.~\ref{fig:teaser}) created by PlotThread.
He mentioned that ``the diverse visual forms of the storylines can arouse viewer's emotions, which I never expect from the optimization-based results.''
Due to the various visual elements proposed in the design space~\cite{Chen2017explore}, we believe that we can further improve PlotThread and expand its applications.
He suggested that ``it can be helpful if the AI agent can guide users when they have no clues on how to start the design of storylines.''
This comment involves the trade-off between AI-driven and AI-guide systems where users or agents start the design process, respectively.
To balance the two sides adequately, we plan to extend our RL framework to enable the agent to generate storylines from the input stories directly.
\section{Discussion}
We discuss the implications and limitations of PlotThread as follows:

\textbf{Implications.}
Our work has several important implications.
First, we develop PlotThread that facilitates the easy creation of storyline visualizations.
Despite that existing tools~\cite{Liu2013,Tang2018a} have incorporated human creativity into the optimization models, they require users to have a deep understanding of the automatic generation process of storyline visualizations.
Thus, non-expert users are usually limited in fully expressing their ideas and design talents when designing the layouts of storylines.
Due to the assistance of the AI agent, PlotThread enables users to design storyline layouts flexibly without considering the aesthetic goals (\textbf{G1} to \textbf{G3}).
The AI agent can resemble the user-specified layouts while preserving the aesthetic quality as much as possible.
Thus, we believe PlotThread can serve numerous amateur users, which reflects the idea of ``visualization for the mass.''

Second, to the best of our knowledge, we are the first to apply reinforcement learning to the design of storyline visualizations.
Despite recent studies indicate that machine learning techniques can be successfully applied to the design of data visualizations (e.g., graphs~\cite{Kwon2019a,Wang2019a} and charts~\cite{Poco2017a}), it is still unknown whether storylines can be produced using machine learning approaches.
To answer this issue, we employ reinforcement learning that formulates the design of storyline visualizations as a long-term delayed reward problem.
The agent is trained to learn how designers typically ``refine'' initial storyline layouts to provide users possible suggestions of effective storyline layouts that follow the aesthetic goals.
Our RL framework can inspire promising research frontiers in the field of visualization design.
For example, researchers could first decompose a complicated design task into a set of design actions and then employ reinforcement learning to predict possible combinations of actions to construct data visualizations.

Third, we propose a mixed-initiative approach that incorporates predictive models and user feedbacks into interactive applications where users initiate and exploit the design task while computational agents explore the design space.
Compared with a typical computer-assisted tool (e.g., iStoryline~\cite{Tang2018a}), PlotThread intends to achieve a better trade-off between human creative work and automation by providing intelligence-level (not tool-level) assistance.
Despite that the optimization-based approaches~\cite{Tanahashi2012,Liu2013,Tanahashi2015} have been improved significantly, we argue that it is necessary to integrate human intelligence into the design of storylines.
Sitting in opposition to a perspective of pure automation, PlotThread provides a successful example where computational agents and people are seamlessly integrated to work on a shared problem, which can inspire the development of future visualization tools.

\textbf{Limitations.}
Our work has several limitations.
First, while there are various design tools to support the design of expressive storylines, PlotThread could be further improved to increase the artistry of the storyline visualizations.
For instance, more diverse sketch styles could be employed to enrich the narration of storylines.
Thus, we plan to develop more design tools to support the creative design of storylines.
Second, even though the time efficiency of the AI agent is acceptable during the authoring process (see Fig.~\ref{fig:experiment-quantitative}), it could be further improved to support more tightly collaborative designs.
As a proof of concept, we have implemented PlotThread on a personal laptop, and we plan to improve the time performance of the AI agent via GPUs and parallel programming.
Third, it is labor-intensive for users to prepare story scripts~\cite{Tang2018a} that are necessary input for the storyline renderer~\cite{Tang2018a}.
To alleviate users' burden, we also plan to enable users to create storyline visualizations from scratch and investigate how to improve the collaborative design workflow progressively.

\section{Conclusion}
In this research, we develop PlotThread, a mixed-initiative authoring tool that seamlessly integrates computational agents and people to facilitate the easy design of storyline visualizations.
The agent is designed to help users explore the design space~\cite{Tang2018a} efficiently by providing a set of suggestive layouts, which can also inspire lateral thinking~\cite{De2010a}.
To develop such an intelligent agent, we formulate the design of storyline layouts as a reinforcement learning (\textbf{RL}) problem where the agent is trained to ``refine'' storyline layouts based on user-shared interactions.
Moreover, we propose a novel framework and generate a collection of well-optimized storylines to address the two major challenges, namely \textit{model architecture} and \textit{model training}, raised by applying RL on designing storylines.
We evaluate the effectiveness of our framework using qualitative and quantitative experiments and demonstrate the usage of PlotThread through a group of use cases.
As future work, we plan to improve the time efficiency of the agent by employing parallel computing and extend the design tools of PlotThread to support more creative designs.

\acknowledgments{
The work was supported by the joint Sino-German program of NSFC (61761136020), National Key R\&D Program of China (2018YFB1004300), NSFC-Zhejiang Joint Fund for the Integration of Industrialization and Informatization (U1609217), Zhejiang Provincial Natural Science Foundation (LR18F020001) and the 100 Talents Program of Zhejiang University.
This project was also partially funded by Microsoft Research Asia.
Lingyun Yu is supported by XJTLU Research Development Funding RDF-19-02-11.
Parts of this work were funded by German Science Foundation (DFG) as part of the project 'VAOST' (392087235).
}

\end{spacing}


\bibliographystyle{abbrv}

\newpage
\bibliography{reference}
\end{document}